\documentclass{article}

\usepackage{PRIMEarxiv}

\usepackage[utf8]{inputenc} 
\usepackage[T1]{fontenc}    
\usepackage{hyperref}       
\usepackage{url}            
\usepackage{booktabs}       
\usepackage{amsfonts}       
\usepackage{nicefrac}       
\usepackage{microtype}      
\usepackage{lipsum}
\usepackage{fancyhdr}       
\usepackage{graphicx}       
\graphicspath{{media/}}     

\usepackage{cite}
\usepackage{amsmath,amssymb,amsfonts}
\usepackage{algorithmic}
\usepackage{graphicx}
\usepackage{textcomp}
\usepackage{xcolor}
\usepackage{subfig}
\usepackage{multirow}
\usepackage{listings}
\usepackage{tikz}
\usepackage{subcaption}
\usepackage{array}
\usepackage{xcolor}
\usepackage{colortbl}
\usepackage{adjustbox}
\usepackage{algorithm2e}
\usepackage{tabularx}
\usepackage{url}
\usepackage[most]{tcolorbox}
\usepackage{hyperref}
\usepackage{balance}

\title{Dynamic Deception: When Pedestrians Team Up to Fool Autonomous Cars
}

\author{
  Masoud Jamshidiyan Tehrani \\
  Università della Svizzera italiana \\
  Lugano, Switzerland \\
  \texttt{masoud.jamshidiyantehrani@usi.ch} \\
   \And
  Marco Gabriel \\
  Università della Svizzera italiana \\
  Lugano, Switzerland \\
  \texttt{marco.gabriel@usi.ch} \\
   \And
   Jinhan Kim \\
  Università della Svizzera italiana \\
  Lugano, Switzerland \\
  \texttt{jinhan.kim@usi.ch} \\
   \And
   Paolo Tonella \\
  Università della Svizzera italiana \\
  Lugano, Switzerland \\
  \texttt{paolo.tonella@usi.ch}
}

\begin{document}
\maketitle

\begin{abstract}
Many adversarial attacks on autonomous-driving perception models fail to cause system-level failures once deployed in a full driving stack. The main reason for such ineffectiveness is that once deployed in a system (e.g., within a simulator), attacks tend to be spatially or temporally short-lived, due to the vehicle's dynamics, hence rarely influencing the vehicle behaviour. In this paper, we address both limitations by introducing a system-level attack in which multiple dynamic elements (e.g., two pedestrians) carry adversarial patches (e.g., on cloths) and jointly amplify their effect through coordination and motion.
 We evaluate our attacks in the CARLA simulator using a state-of-the-art autonomous driving agent. At the system level, single-pedestrian attacks fail in all runs (out of 10), while dynamic collusion by two pedestrians induces full vehicle stops in up to 50\% of runs, with static collusion yielding no successful attack at all. These results show that system-level failures arise only when adversarial signals persist over time and are amplified through coordinated actors, exposing a gap between model-level robustness and end-to-end safety.
\end{abstract}


\section{Introduction}
\label{sec:intro}



Autonomous Cars (ACs) increasingly rely on Deep Learning (DL) based perception modules to interpret their surroundings and make safety-critical decisions~\cite{grigorescu2020survey}. While DL models have enabled impressive advances in autonomous driving, their tight coupling with vehicle control has also exposed critical vulnerabilities~\cite{Eykholt_2018_CVPR}. Over the past years, a large body of work has demonstrated that such perception components can be fooled into misclassifying objects, hallucinating traffic signs, or ignoring critical obstacles~\cite{Thys_2019_CVPR_Workshops}.

While such attacks have been successful, much of the existing literature has focused on \textit{model-level} vulnerabilities and \textit{static} attack settings~\cite{Eykholt_2018_CVPR}. Typical attacks manipulate a single frame or a fixed element of the environment, such as billboards or traffic signs~\cite{zhou2020deepbillboard}. Their success is often measured in terms of misclassification rates of the DL component in the AC, without necessarily examining whether these errors propagate to the system-level, resulting in faulty driving behaviors~\cite{hu2022pass}. As prior work has shown~\cite{qian2025review,tehrani2025taxonomy}, model-level failures do not automatically translate into system-level failures: the software stack behind an AC integrates perception over time, fuses multiple sensors, and applies behavioral logic that can isolate the failures from the low-level. As a result, attacks that appear powerful in isolation might be ineffective once embedded in a driving system.

Moreover, in a real traffic context, perception is not static. ACs operate in dynamic environments populated by other actors, such as cars, cyclists, and pedestrians, all of which move, interact, and continuously enter and leave the car's field of view. Among these actors, pedestrians play a prominent role as they are highly salient to the perception systems, which often prioritize them for safety reasons. Although previous work has used them within threat models of adversarial attacks~\cite{tehrani2025taxonomy}, they are treated as passive background elements or as a detection target, rather than as active carriers of attack.

In this paper, we present a system-level attack that exploits dynamic actors, such as pedestrians, as carriers of adversarial patches. To make the attack \textit{stealthy}, we embed an adversarial stop-sign pattern into pedestrians’ clothing. However, our analysis showed that the physical size of a cloth, such as a T-shirt, significantly limits the strength of the attack, as it is much smaller than a standard traffic sign. This limitation can be overcome through the \textit{collusion} of multiple pedestrians (two, in our study), who collectively form a larger effective patch by spatially aligning partial patterns from the car’s camera viewpoint. Crucially, we further show that pedestrian motion is a decisive factor. Static pedestrians, even when colluding, do not induce a system-level failure. In contrast, when pedestrians move alongside the car and remain within its field of view, perception errors persist over time and mislead the AC’s perception model, ultimately resulting in a misbehavior, such as a full stop of the car.

We conduct an extensive empirical evaluation in the CARLA simulator~\cite{Dosovitskiy17} with state-of-the-art autonomous driving agents. Our experiments systematically compare a single-pedestrian attack to a collusion attack, and static scenarios to dynamic scenarios.  Results show that single-pedestrian attacks consistently fail to stop the car, even though they bias the perception model toward stop signs. Static collusion attacks also fail to propagate to the car behavior. Only the dynamic collusion attack, in which multiple pedestrians move in coordination with the car, reliably causes the AC to execute a full stop. These findings demonstrate that dynamics and persistence are essential for turning model-level failures into system-level failures.

In summary, this paper makes the following technical contributions:

\begin{itemize}
    \item We introduce a \textit{stealthy} system-level attack technique that uses pedestrians as carriers of adversarial patches.

    \item We propose \textit{collusion pedestrian attacks}, in which colluding pedestrians jointly form a larger adversarial patch, overcoming patch size constraints. 

    \item We propose \textit{dynamic pedestrian attacks}, in which pedestrians walk in front of the AC, continuously inducing the adversarial effect on the autonomous agent.

    \item We provide an extensive experimental evaluation in the CARLA simulator against a state-of-the-art autonomous driving agent.
\end{itemize}
\section{Background}
\label{sec:background}

\subsection{Autonomous Cars}

According to Faisal et al.~\cite{faisal2019understanding}, an AC can only be considered truly autonomous when it is capable of performing all dynamic driving tasks seamlessly across a wide variety of environments. Vehicles with automated features exhibit varying levels of autonomy. In the case of ACs, five distinct levels have been defined~\cite{SAE2021J3016}, ranging from basic driver assistance to full driving automation, the latter being the level Faisal et al. refer to. 
Modern ACs typically coordinate multiple modules to achieve autonomous driving with enhanced safety and security~\cite{levinson2011towards}. These modules include several \textit{perception modules}, which process raw data from sensors, such as cameras, LiDAR, and radar, to detect and classify objects, identify lanes, and understand the environment. These modules form the vehicle’s situational awareness foundation, obtained by extracting and fusing meaningful features necessary for safe navigation.


\subsection{Adversarial Attacks}
DL models have demonstrated remarkable performance across a wide range of application domains. However, they remain vulnerable to \textit{adversarial inputs}: intentionally manipulated inputs, crafted to mislead detection systems at inference time. Studies have shown that deep neural networks used for object recognition or classification can be deceived by subtle alterations to input images, modifications that are virtually imperceptible to the human eye~\cite{biggio2018wild}. An \textit{adversarial attack} refers to a method in which an attacker subtly manipulates input data to cause a DL model to make incorrect predictions. These attacks are particularly concerning in computer vision systems, where small, often imperceptible changes to an image can cause a DL model to misclassify it~\cite{goodfellow2014explaining}.
    
The following are key properties of adversarial attacks: universality, targeted mis-detection, locality, stealthiness and transferability. An attack perturbation is \textit{universal}, if it can be added to any input sample from a given dataset to cause a trained DL model to mis-classify the input with high probability. An attack is \textit{targeted} if it triggers the detection of a specific target class chosen by the attacker (e.g., a stop sign). Attack perturbations are \textit{localized} if they operate on a small area of the input image (e.g., the T-shirt of an adversarial pedestrian). \textit{Stealthiness} makes the attack difficult to detect for human observers or automated detectors. Finally, \textit{transferability} ensures the possibility of applying the attack to different DL models.

To generate adversarial attacks, the attacker can have different knowledge of the model and the system under attack~\cite{tehrani2025taxonomy}. Attacker's knowledge is categorized into white-, black-, or gray-box settings depending on the level of visibility into the system or model. In a Model-level White-Box scenario, the attacker has full access to the code used to create the model and its training datasets, while a Model-level Black-Box scenario restricts knowledge to only the input and output vectors. At the broader system level, a White-Box setting grants unrestricted access to the entire infrastructure, including source code, hardware specs, DL models, and simulators, whereas a System-level Black-Box limits the attacker to observing sensor inputs and actuator outputs with no internal visibility. Finally, a System-level Gray-Box represents a middle ground where the attacker possesses partial insights, such as access to the code of specific components, without having full transparency of the entire system.
\section{Related Work}
\label{sec:relatedwork}

Adversarial T-shirts, as introduced by Xu et al.~\cite{xu2020adversarial}, represent a model-level attack in which patches printed on clothing are used to evade person detection. The method is designed to remain robust against the non-rigid transformations that occur as a person moves. This robustness is achieved by applying thin-plate spline transformations to the patch during training.

Closely related to our own work, the approach by Wang et al.~\cite{wang2023driving} applies an adversarial patch to the clothing texture of pedestrians in order to compromise the Transfuser~\cite{Chitta2023PAMI} autonomous driving agent in the CARLA simulator~\cite{Dosovitskiy17}. The goal of their attack is to increase the vehicle’s speed to the point where it collides with stationary pedestrians. Their method requires complex computational steps to generate an effective patch and apply it to the pedestrian model. The proposed method has several limitations that prevent its general applicability: (1) to provide an artificially large patch to the vision system, the pedestrian is positioned in the middle of the road, facing the car, which is quite unrealistic; (2) the pedestrian does not move as the car approaches it, which artificially inflates the effectiveness of the patch. Our experiments show that attacks carried by a single static pedestrian at the roadside are not effective enough to propagate to system-level failures.

Most system-level attacks on autonomous vehicles rely on static elements in the environment, such as billboards~\cite{von2023deepmaneuver} or the road itself~\cite{sato2021dirty, nassi2020phantom}, to deliver the adversarial signal. Hanfeld et al.~\cite{hanfeld2023kidnapping}, however, introduced an adversarial drone that moves dynamically with the ego drone to maintain the effectiveness of the attack. They propose a method capable of hijacking multi-rotor drones using flying adversarial patches, i.e., small images dynamically positioned in the environment. As a result, the target drone deviates from its intended trajectory, loses track of the human target, and instead follows the adversarial patch. While this approach shares the dynamism of the attack with us, it does not take \textit{stealthiness} into account at all: patches are large printings held by an adversarial drone, which would immediately look suspicious and anomalous to any manual or automated attack detection system. We instead use multiple pedestrians' T-shirts, which do not even trigger the attack if considered in isolation.

Overall, our work is the first to propose collusive, dynamic pedestrian attacks against ACs that maximize the attack effectiveness while maintaining a high degree of stealthiness.

\section{Approach}
\label{sec:approach}

\begin{figure}[!h]
    \centering
    \subfloat[Original]{\includegraphics[width=0.28\columnwidth]{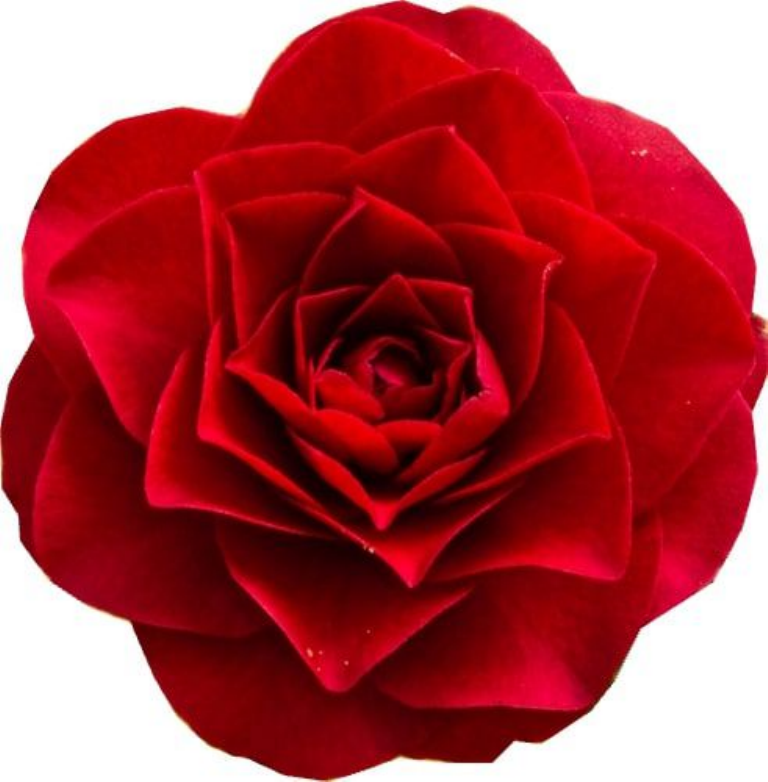}}
    \label{fig:camellia_flower}
    \hspace{0.04\columnwidth}
    \subfloat[Patched]{\includegraphics[width=0.28\columnwidth]{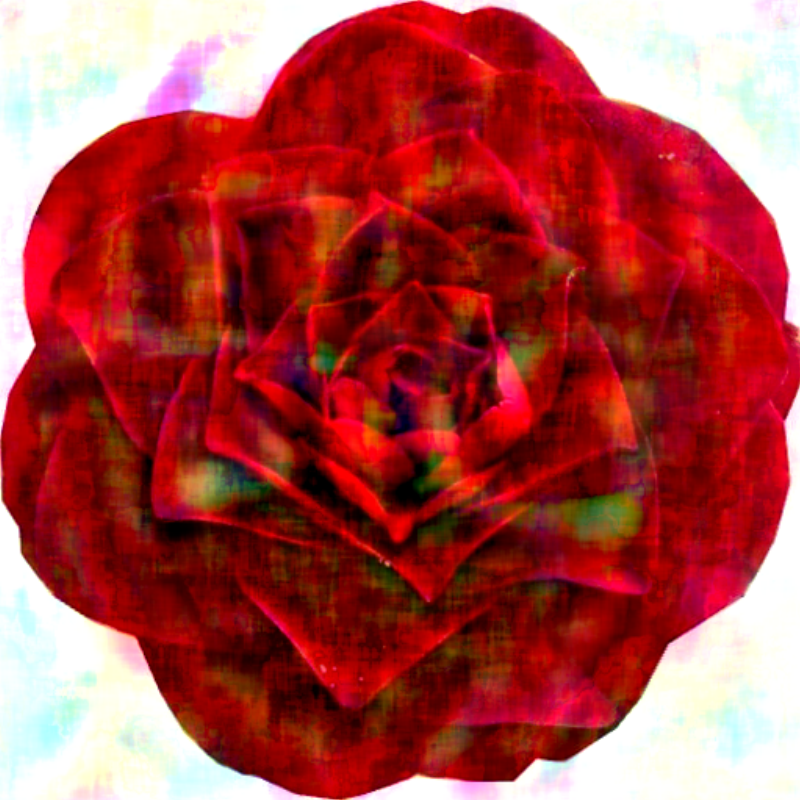}}
    \label{fig:adversarial_camellia}
    \caption{Camellia flower next to the patched camellia}
    \label{fig:camellia_compare}
\end{figure}

Our objective is to use pedestrians to implement a stealthy attack that is activated only when multiple pedestrians cooperate (collude) to expose the attack dynamically to the victim AC.  To implement this attack, we print attack patches on pedestrians' T-shirts and let pedestrians walk along the pavement in a coordinated way. We refer to these pedestrians as \textit{adversarial pedestrians}. Our base image for adversarial patch generation is a red camellia flower, chosen for its similarity in color and general shape to the stop signs used in the CARLA simulator~\cite{Dosovitskiy17}. We then patch the camellia image with an adversarial stop-sign perturbation specifically crafted to remain visually concealed within the flower. Fig.~\ref{fig:camellia_compare} shows the camellia image next to the patched camellia image.

The adversarial image must remain stealthy to human observers, while causing ACs to perceive it as a real stop sign. To accomplish this, we perform a black-box model-level attack, as limited access to the internal DL models of ACs reflects realistic deployment conditions. In such a setting, the adversarial patch is trained on a surrogate model, which is another DL model with similar functionality to the target one. Prior research has shown that adversarial samples can transfer across models trained for the same task~\cite{papernot_2016_transferability}. Then, we design dynamic deployment scenarios that enhance the likelihood of a successful attack at the system level. We consider both single and collusion attacks, using one and two pedestrians, respectively. We then exploit the pedestrians’ ability to walk to add dynamism to the attack.


\subsection{Patch Algorithm and Frameworks}
Leveraging transferability, we employ a gradient-based method on a surrogate model to generate adversarial patches. Gradient-based approaches are widely used due to their effectiveness in producing optimal, universal, and physically realizable perturbations~\cite{moosavi2017universal}. To meet our objectives, we require an attack algorithm capable of producing universal, targeted perturbations restricted to a localized area, namely, a T-shirt patch. Additionally,  stealthiness and transferability of the adversarial patch are essential.

The adversarial patch generation algorithm proposed by Brown et al.~\cite{brown2018adversarialpatch} is well-suited to our use case, as it satisfies all of our requirements; therefore, we adopt it in this work. The Adversarial Robustness Toolbox (ART)~\cite{art2018} provides a direct implementation of this algorithm, along with a wide range of other adversarial methods. Given its widespread adoption and active development, we select ART as the framework for our attack.

\subsection{Dataset Collection}
To train our adversarial patch, we require a dataset of CARLA images captured across different towns. Since our goal is to mislead the system into detecting pedestrians as stop signs, the dataset must include pedestrians walking on sidewalks, closely reflecting our attack scenario. 

To ensure compatibility with the autonomous agents in the CARLA simulator, the dataset should resemble the images typically used to train the agents as closely as possible. For this reason, we adopt Bench2Drive~\cite{bench2drive}, the benchmark dataset used for the Transfuser++ agent~\cite{hidden_biases_tfpp} as our data collection reference. Following Bench2Drive’s setup, we mounted a camera on top of a vehicle and configured its attributes accordingly. These attributes include field of view (FOV), resolution, and other parameters affecting image sharpness, such as blur amount, motion blur, shutter speed, and aperture (f-stop). The exact values are provided in the replication package. Unlike Bench2Drive, however, we consistently use the \texttt{ClearNoon} weather setting, as it sufficiently demonstrates that our attacks remain effective under favorable lighting and visibility conditions.

In total, we collected 1,517 images across 28 scenarios spanning 5 towns, totaling approximately 4 GB. The distribution per town is as follows: Town01 (501 images),  Town02 (203), Town03 (170), Town04 (334), and Town05 (309). The dataset is divided into 1,213 training images (80\%) and 304 validation images (20\%). This dataset collection technique places our attack in the category of \textit{system-level gray-box attacks}, as we possess only partial information about the specifications of the hardware used by the agents. 

\subsection{Adversarial Patch Generation}
To generate our model-level black-box attack, we need a surrogate model to train the adversarial patch. We select the YOLOv5~\cite{redmon2016you} model family due to its widespread usage, its speed and low memory requirements, which keep training times feasible on consumer hardware.

We load and preprocess the dataset, which involves removing any non-RGB channels. We then pass all images through the model to obtain predictions on benign images. 

\begin{figure*}[t]
  \centering
  \subfloat[Adversarial patch]{%
    \includegraphics[width=0.3\textwidth]{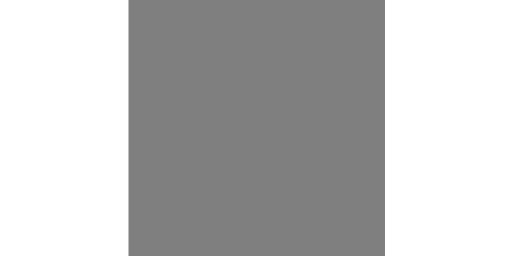}\label{fig:patch_gen_proc_a}}
  \hspace{0.01\textwidth}
  \subfloat[Benign image]{%
    \includegraphics[width=0.3\textwidth]{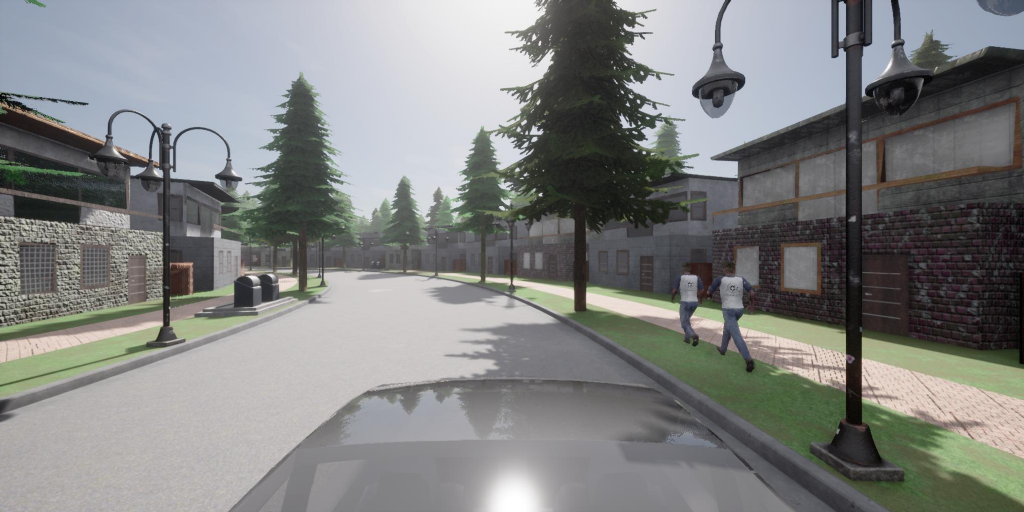}\label{fig:patch_gen_proc_b}}
  \hspace{0.01\textwidth}
  \subfloat[Patch applied]{%
    \includegraphics[width=0.3\textwidth]{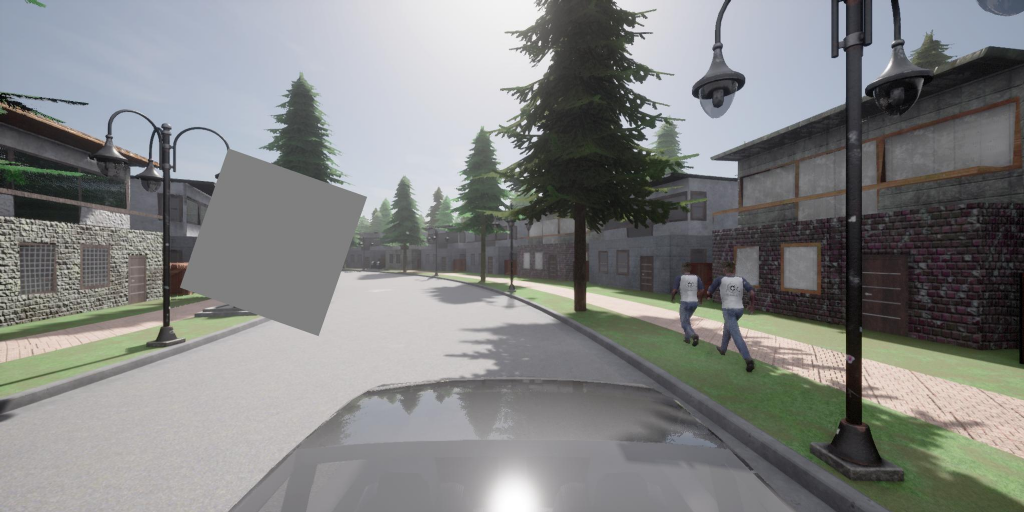}\label{fig:patch_gen_proc_c}}
  \hspace{0.01\textwidth}
  \hfill
  \subfloat[Detections]{%
    \includegraphics[width=0.3\textwidth]{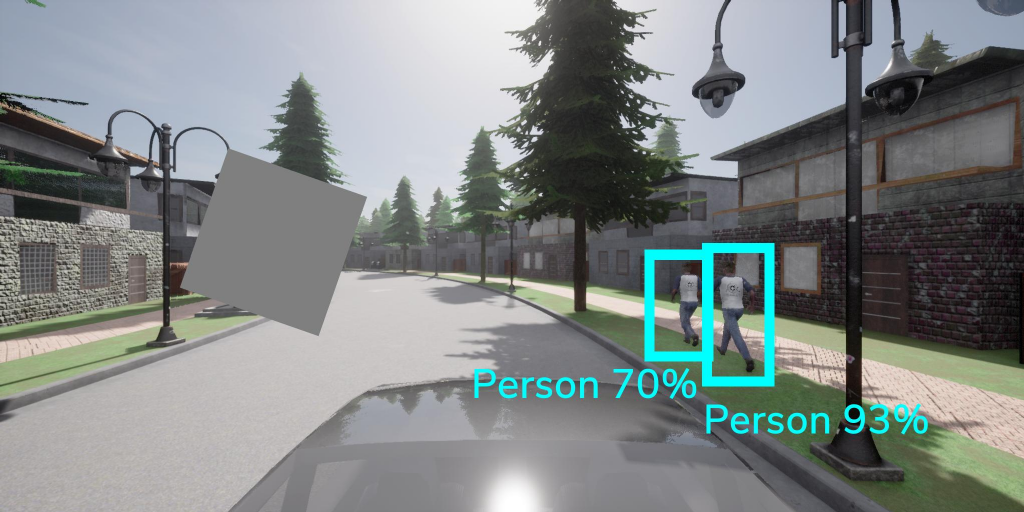}\label{fig:patch_gen_proc_d}}
  \hspace{0.01\textwidth}
  \subfloat[Target detections]{%
    \includegraphics[width=0.3\textwidth]{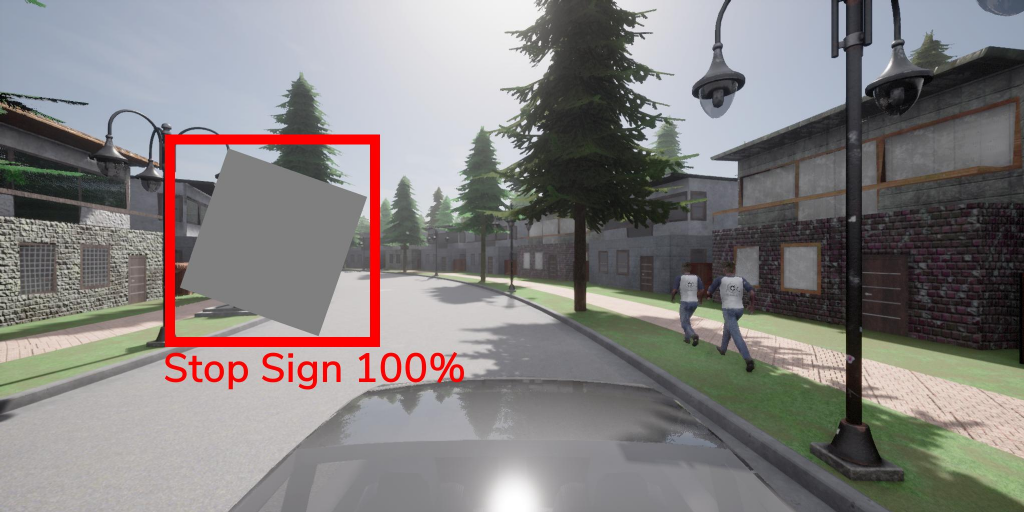}\label{fig:patch_gen_proc_e}}
  \hspace{0.01\textwidth}
  \subfloat[Loss basis]{%
    \includegraphics[width=0.3\textwidth]{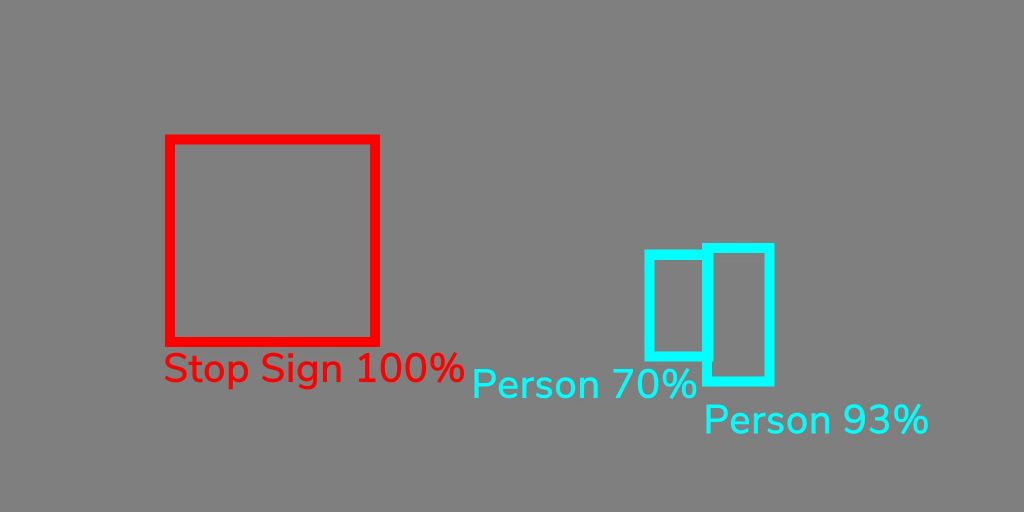}\label{fig:patch_gen_proc_f}}

  \caption{Conceptual steps of the patch generation process}
  \label{fig:patch_gen_proc}
\end{figure*}

Generating an adversarial patch is an iterative process (Fig.~\ref{fig:patch_gen_proc}), beginning with a plain gray patch or a specified base image (Fig.~\ref{fig:patch_gen_proc_a}). This patch is subsequently overlaid on a training image (Fig.~\ref{fig:patch_gen_proc_b}) and subjected to random transformations, including scaling, translation, rotation, and distortion (Fig.~\ref{fig:patch_gen_proc_c}). The resulting \textit{patched} image is processed by the victim object detection model, producing predictions on bounding boxes, classes, and confidence scores (Fig.~\ref{fig:patch_gen_proc_d}). These predictions differ from the adversary’s desired outcome, e.g., detecting a stop sign at the patch location with maximum confidence (Fig.~\ref{fig:patch_gen_proc_e}). A loss is then calculated by comparing these predictions against the adversary’s desired detections, and backpropagated through the model to update the patch pixel-wise using gradient descent (Fig.~\ref{fig:patch_gen_proc_f}).

A crucial part of this process is the transformation applied to the patch. The underlying principle, known as Expectation over Transformation (EoT)~\cite{objects}, ensures the patch’s universality~\cite{brown2018adversarialpatch}. Formally, this is defined in Equation~\ref{eq:EOT}, where the final patch $\hat{p}$ is trained to maximize the expected probability of the adversary’s desired prediction $y$ under random transformations $t$ applied to a training image $x$. 

\begin{equation}
\begin{aligned}
\hat{p} = \arg \max_p \mathbb{E}_{x \sim X, t \sim T} [P(y \mid A(x, t, p))]
\end{aligned}
\label{eq:EOT}
\end{equation}

\begin{figure*}[t]
    \centering
    \subfloat[Patch after 1 training step]{\includegraphics[width=0.30\textwidth]{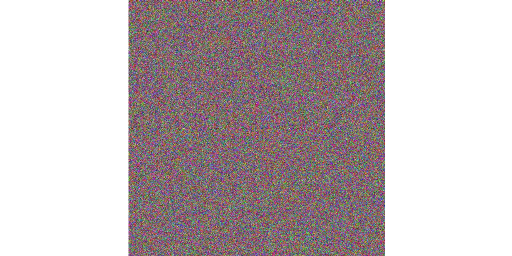}\label{fig:patch_gen_proc_2a}}
    \hspace{0.01\textwidth}
    \subfloat[Detections on patched image]{\includegraphics[width=0.30\textwidth]{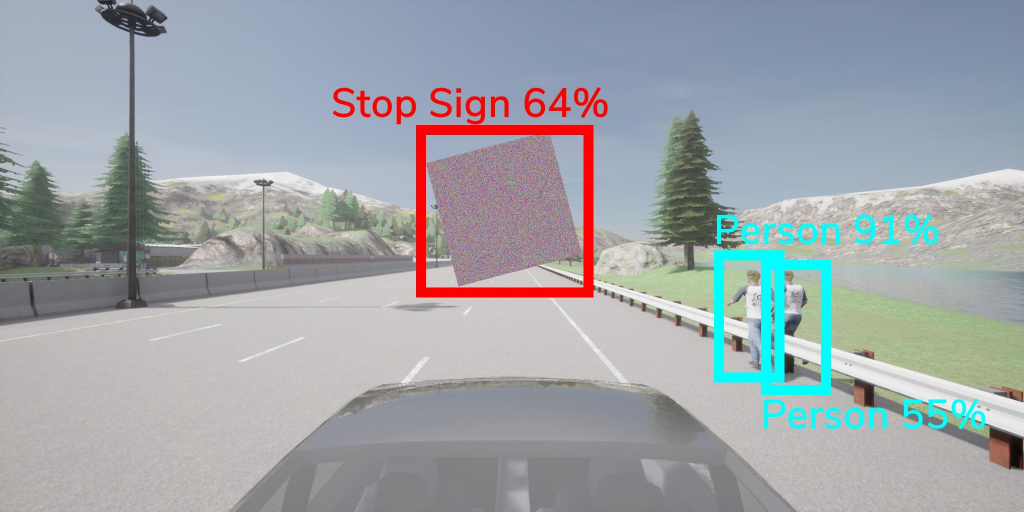}\label{fig:patch_gen_proc_2b}}
    \hspace{0.01\textwidth}
    \subfloat[Loss basis with desired detection in purple]{\includegraphics[width=0.30\textwidth]{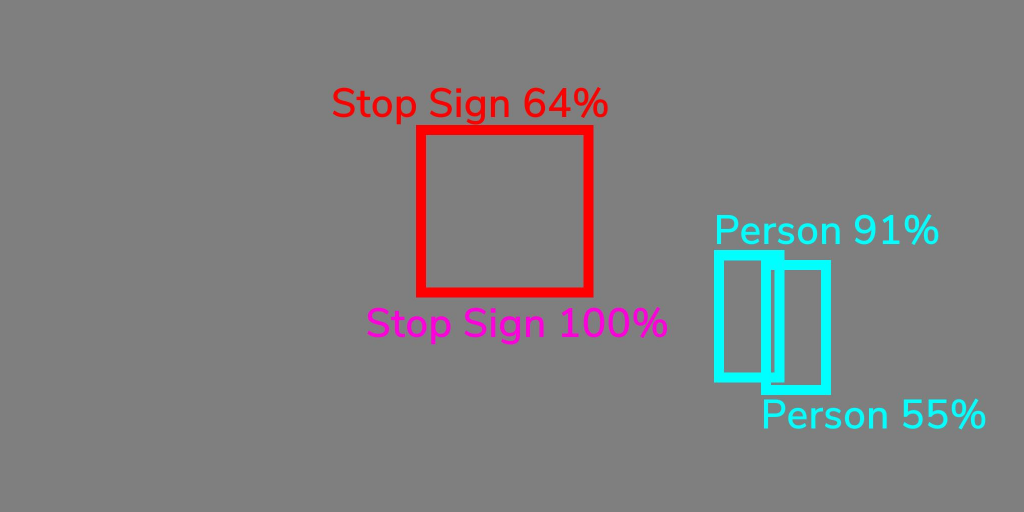}\label{fig:patch_gen_proc_2c}}
    
    \caption{Selected steps of the patch generation process during the second iteration }
    \label{fig:patch_gen_proc2}
\end{figure*}

In subsequent iterations, the patch is applied to new training images with new transformations. As training progresses, the patch increasingly induces targeted (e.g., stop-sign) detections at its location, reducing the loss. Fig.~\ref{fig:patch_gen_proc2} illustrates the process after the first training step.

\subsection{Disguise}

For the base image used in the patch generation process, we initially experimented with various images, including a mountain scene, but these did not yield good model-level results. We therefore selected a red camellia flower, whose color and structure more closely resemble a stop sign, leading to a stronger and more reliable attack.

A stealthy adversarial patch should appear natural and describable to humans, without resembling the attack’s target class. Such an attack is called a \textit{disguise}~\cite{disguise}. To achieve this, we incorporate a disguise loss term into our optimization objective, encouraging the patch to resemble a chosen disguise image while preserving adversarial effectiveness. 
We use the $L_2$ norm to measure similarity, as shown in Equation~\ref{eq:disguise}. Here, $L_{disguise}$ is the Euclidean distance between the patch and the disguise image. The total loss, $L_{total}$, is the sum of the adversarial loss $L$ (known from the literature~\cite{chow2020tog} and consisting of three terms, namely, classification, confidence, and box location loss) and the weighted disguise loss, controlled by hyperparameter $k_{disguise}$. 

\begin{equation}
\begin{aligned}
L_{disguise} = ||Patch - Disguise||_2 \\
L_{total} = L + k_{disguise} \cdot L_{disguise}
\end{aligned}
\label{eq:disguise}
\end{equation}

\begin{figure}
    \centering
    \includegraphics[width=0.65\columnwidth]{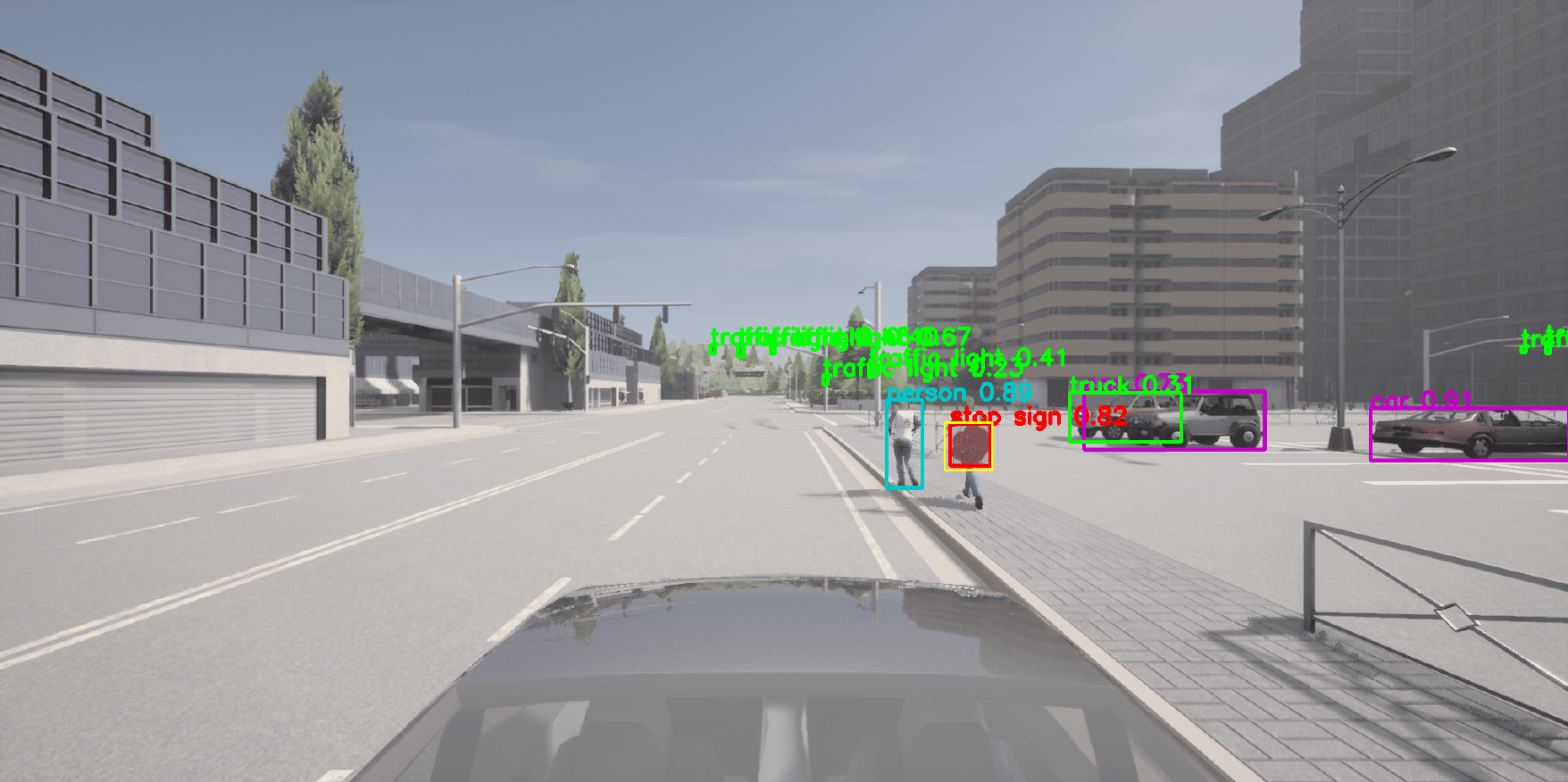}
    \caption[Patch generation evaluation]{Example of a patch evaluation}
    \label{fig:patch_eval}
\end{figure}

To evaluate the effectiveness of our trained patches, we programmatically apply them to randomly selected unseen images from the test set. The application process, as illustrated in Fig.~\ref{fig:patch_eval}, mirrors the training setup, with transformations (e.g., scaling, rotation, translation, and distortion) within the same parameter ranges. At inference time, such transformations are applied to check for the robustness of the patch, anticipating what could happen at the system-level when, e.g., the patch moves with the walking pedestrians, or the AC camera's perspective varies, depending on the distance.

To avoid confounding factors, a detection is only counted as successful if the predicted stop sign overlaps with the patch's location in the image. This check ensures that only detections directly caused by the patch are considered, while eliminating positives arising from other sources, for example, real traffic signs or unrelated background objects. For instance, if a billboard in the background were misclassified as a stop sign, without filtering, such detection would artificially inflate the attack success rate.


\subsection{Single and Collusion Attacks}

After generating the camellia patch and before applying it to the pedestrians’ T-shirts, we first test the patch by replacing the default stop-sign texture on a stop-sign pole with the patch texture. This confirmed that the patch functions correctly, as the vehicle stopped in a manner comparable to when the original stop-sign texture was present. 

Then, we scaled the pedestrian model to an (unrealistically) larger size such that the camellia patch matched the dimensions of a standard stop sign. Under these conditions, the attack succeeded, while it did not without rescaling, hence confirming that the \textit{patch size} is a critical limiting factor: the patch must approximate the size of the default stop sign in the CARLA simulator to be effective. This insight motivated the use of two pedestrians in a \textit{collusion attack}, to increase the combined visible patch area, thereby improving the attack’s effectiveness.

Accordingly, we apply the camellia patch under two attack settings: \textit{single} and \textit{collusion}. In the single attack, the entire patch is printed on the T-shirt of a single pedestrian. In the collusion attack, the patch is divided into two halves, each printed on the T-shirt of a different pedestrian. The effectiveness of the collusion attack depends critically on the camera viewpoint. Our objective is to increase the perceived patch size by combining the two halves across two pedestrians. Simply placing the pedestrians side by side is ineffective because the camera, positioned slightly to the left, does not capture them as visually adjacent. Instead, we place the left pedestrian slightly behind the right one so that the two patch halves align from the camera’s perspective and appear as a single, larger pattern. This configuration is illustrated in Fig.~\ref{fig:collusion_attack}.

\subsection{Static and Dynamic  Attacks}
Replacing a stop sign with pedestrians is considered a \textit{static pedestrian attack} when the pedestrians remain stationary. However, since pedestrians can walk and adjust their speed, we leverage this ability to introduce a dynamic aspect to our attack. A \textit{dynamic pedestrian attack} refers to scenarios in which the pedestrians move at approximately the same speed as the car, remaining in its field of view for a longer duration and maximizing the visibility of the patch to the car’s camera. This requires a road where the car’s speed is low enough for the pedestrians to tentatively keep pace with the vehicle. In this attack, we position the pedestrians approximately 6 meters behind a possible location for a stop sign (e.g., an intersection) and move them according to the vehicle’s speed, computed as the Euclidean norm of $\langle v_x, v_y, v_z\rangle$, the vehicle’s velocity components along the $x$-, $y$-, and $z$-axes. 
%
%
Using this velocity, we move the pedestrians toward the target intersection, but stop them before they enter the roadway, keeping them at a safe and realistic distance on the pavement.
\section{Experimental Setup}
\label{sec:experimentalsetup}

\subsection{Research Questions}
\label{sec:rq}
To evaluate the effectiveness of the proposed approach, we design an experiment aimed at answering the following research questions:

\noindent\textbf{RQ1 [Single Attack]:} \textit{What is the effectiveness of the single pedestrian attack?} \\
This question seeks to examine whether a single pedestrian can cause the vehicle's perception system to misclassify it as a stop sign, resulting in the temporary stop of vehicle movement. 

\noindent\textbf{RQ2 [Collusion Attack]:} \textit{What is the effectiveness of a collusion attack compared to the single pedestrian attack, and how different is it from its two halves considered separately?} \\
This research question investigates whether splitting an image into two halves and placing each half on two pedestrians can improve the effectiveness of the attack compared to using a single pedestrian. The idea is that two pedestrians standing side by side can cover a larger visual area. Additionally, we examine whether each half of the image is less effective on its own. 

\noindent\textbf{RQ3 [Dynamic Attack]:} \textit{Can dynamism make the attack more effective?} \\
Since pedestrians can easily move, we aim to examine whether having them move as the vehicle approaches, thereby keeping the attack visible for a longer duration, can make the attack more effective.


\subsection{Model- and System-level Evaluations}

Our evaluation is divided into two parts: one is a  \textit{model-level} and the other a \textit{system-level} evaluation. In the model-level evaluation, the adversarial patch is tested directly against the AC’s object detection component, prior to assessing its impact at the system level (i.e., the AC’s actions, such as steering and brake, are ignored). In contrast, a system-level evaluation applies the attack within a simulation or real-world setting to determine whether the model’s failure propagates into a system-level malfunction, in our case, a failure of the AC itself.

At the model-level, we inspect the autonomous agent's classification output to determine whether it detected pedestrians or a stop sign. If the agent reports detecting stop signs more frequently than pedestrians, this indicates that the attack successfully misled the classifier into interpreting the pedestrian as a stop sign.

At the system-level, we examine how the agent’s classifications influence the vehicle’s behavior, specifically, whether they cause the car to adjust its motion and to ultimately come to a complete stop. To evaluate this, we conduct experiments in a simulation environment rather than the real world, as a physical attack would be both costly and potentially hazardous for pedestrians. For this purpose, we use the CARLA simulator~\cite{Dosovitskiy17}, which offers several advantages. Most importantly, CARLA is highly customizable, allowing us to control the environment, configure pedestrians, apply adversarial patches to their clothing, and design tailored test scenarios that closely match our attack requirements.

\begin{figure}[t]
    \centering
    \subfloat[Collusion Attack]{\includegraphics[width=0.4\columnwidth]{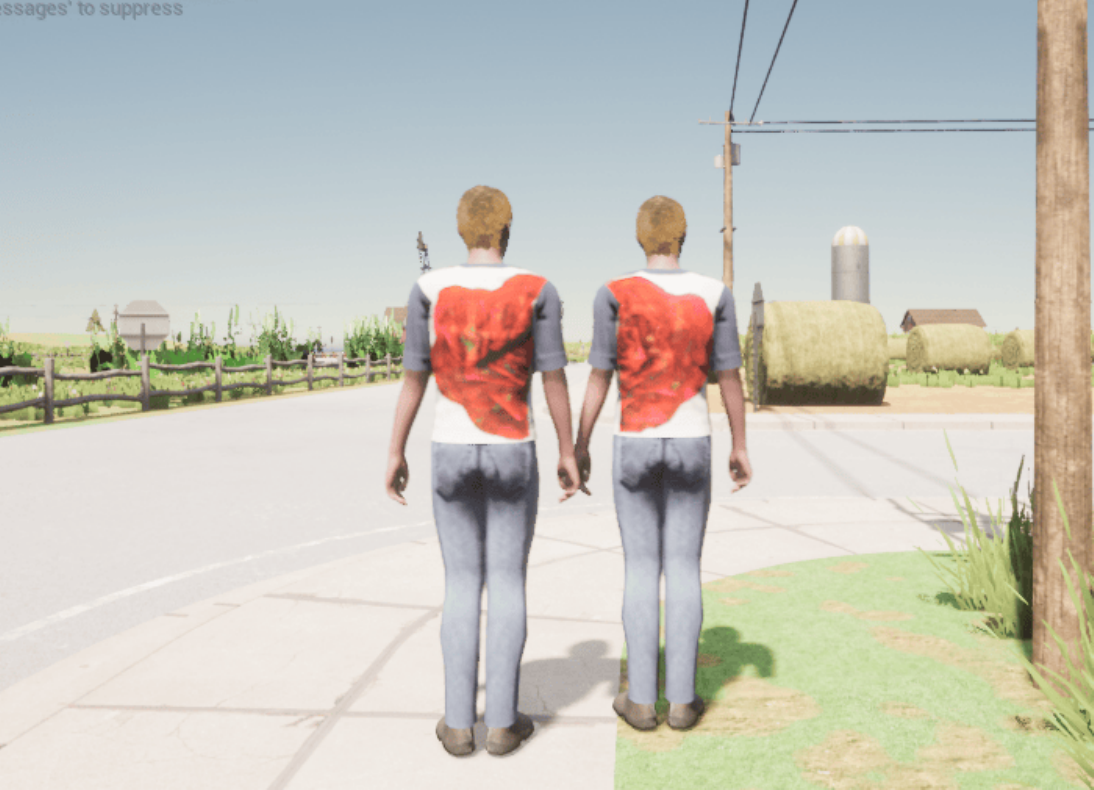}
    \label{fig:collusion_attack}}
    \hspace{0.04\columnwidth}
    \subfloat[Single Attack]{\includegraphics[width=0.4\columnwidth]{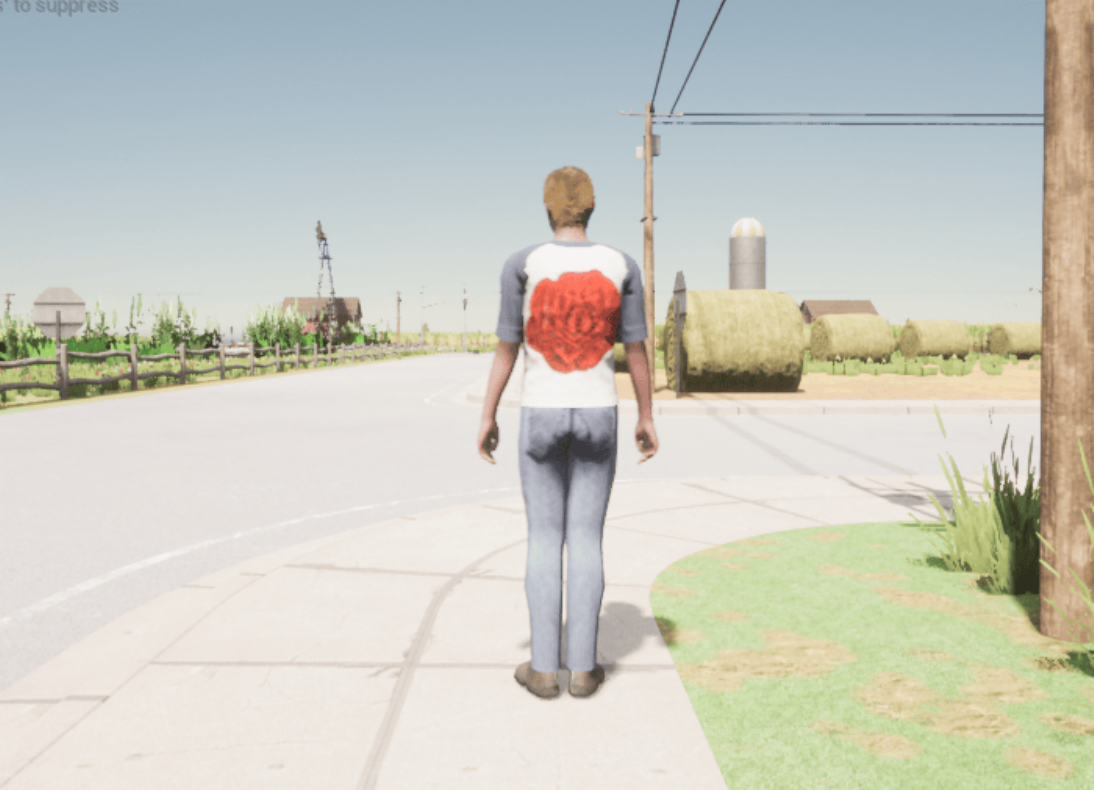}
    \label{fig:single_attack}}
    \caption{Pedestrian shirts in the CARLA simulator with patched camellia image}
    \label{fig:pedestrian_shirts}
\end{figure}

The attacks are implemented in the CARLA simulator (version 9.15.2). To design adversarial scenarios, we require pedestrians wearing adversarial T-shirts. These T-shirts are created by manually modifying the default pedestrian textures to embed our adversarial patch. The modified textures are then applied to the pedestrian models (commonly referred to as \textit{walkers} in the simulation) using the blueprints provided by CARLA. This customization is performed through the Unreal Engine 4 Editor~\cite{unrealengine4}, which is the backbone of the simulation. Fig.~\ref{fig:pedestrian_shirts} shows pedestrians with patches on their T-shirts.

\subsection{Autonomous Agents}
To evaluate our attacks, we require one or more autonomous agents within the CARLA simulator. PCLA~\cite{tehrani2025pcla} offers a straightforward framework for deploying autonomous agents onto the ego vehicle, supporting 27 different high-performing agents from the CARLA Leaderboard challenge~\cite{leaderboard} and enabling seamless switching between them within a unified programming environment.

For our stop-sign attack, we require an agent that reacts to stop signs in the expected way: upon detecting a stop sign, the vehicle must come to a complete stop and ensure that the intersection (or railroad crossing) is safely clear of pedestrians and other vehicles before proceeding~\cite{stipancic2021evaluating}. This means the selected agent should stop fully in the presence of a stop sign, while it should continue driving when no stop sign is present. 

After evaluating all PCLA agents using this criterion, we found that only the Simlingo agent~\cite{Renz2025cvpr} behaves accordingly among all the agents. This may be due to the fact that, in the Leaderboard challenge, the infraction penalty coefficient for running a stop sign is only 0.8, which is significantly lower than many other infraction penalties. As a result, many agents may implicitly learn to deprioritize strict stop-sign compliance. Simlingo uses a language model to interpret visual input and then translates the resulting text output into CARLA control actions to drive the vehicle. It achieved first place in the CARLA Leaderboard 2 SENSORS track. Moreover, since it is open-source, we have access to the language model outputs, which allows us to inspect the results of object detection for a more detailed evaluation.

\subsection{Scenarios}
\begin{figure}[!t]
    \centering
    \subfloat[Intersection with a stop sign]{\includegraphics[width=0.4\columnwidth]{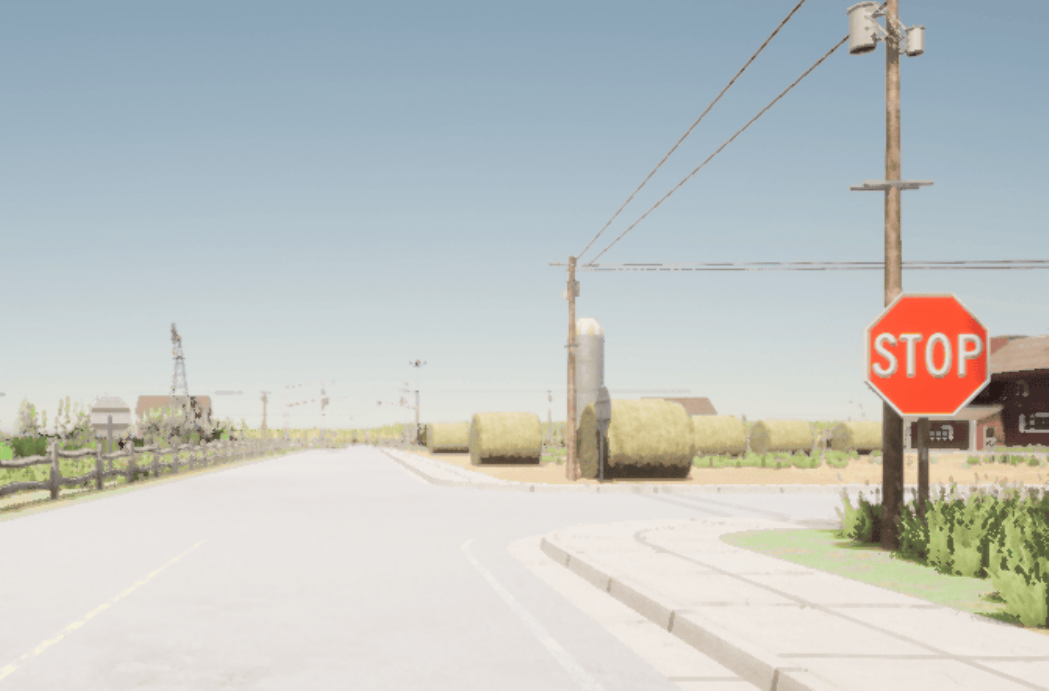}
    \label{fig:intersection_stop_sign}}
    \hspace{0.04\columnwidth}
    \subfloat[Intersection with pedestrians]{\includegraphics[width=0.4\columnwidth]{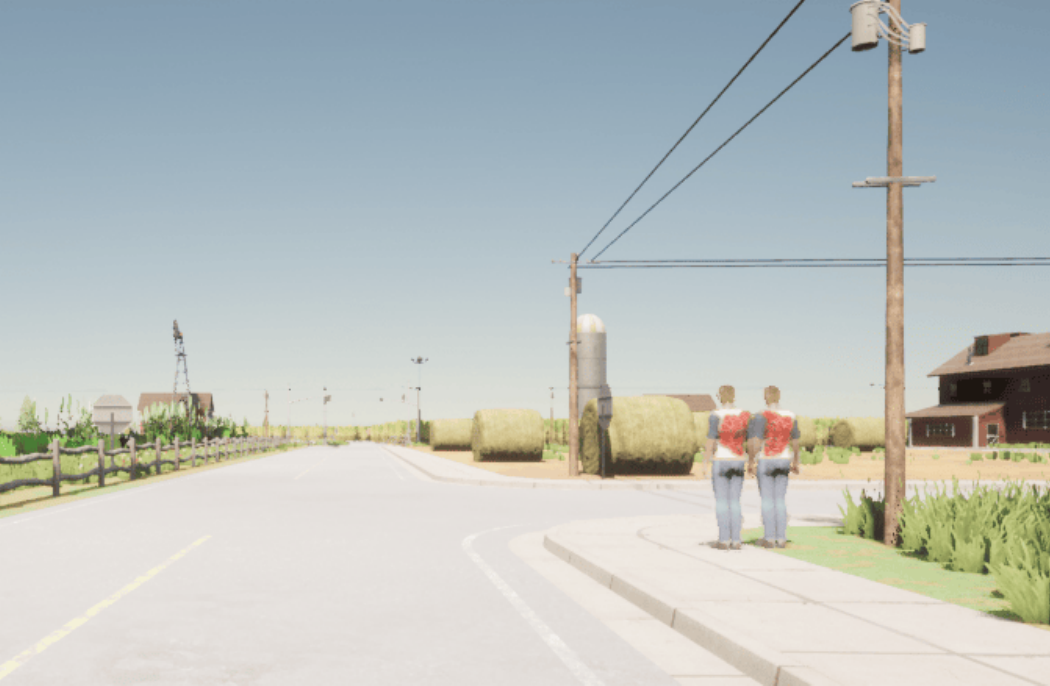}
    \label{fig:intersection_pedestrians}}
    \caption{The selected intersection in Town07 of the CARLA simulator}
    \label{fig:intersection}
\end{figure}

To evaluate our attack, we selected Town07 in the CARLA simulator, focusing on a specific intersection that normally contains a stop sign (Fig.~\ref{fig:intersection_stop_sign}). We removed the stop sign and replaced it with pedestrians, as illustrated in Fig.~\ref{fig:intersection_pedestrians}. 
Some non-autonomous cars were also added in the scenario to make it look more realistic. 

One full run of our experiment is defined as the AC starting from a specified origin point and reaching the designated destination. This means that a run is not determined by time or by a fixed number of frames; instead, its duration may vary depending on the AC’s behavior and actions during the scenario. The starting location is placed far from the pedestrians to minimize noise caused by having them too close to the vehicle at the beginning of the run. The destination point is set beyond the intersection, at a position where neither the pedestrians nor the intersection remains within the vehicle’s line of sight.

\textbf{General Scenarios:}
We define four main scenarios. Two involve using a single pedestrian to carry out the attack (the single attack), and the other two involve using two pedestrians (the collusion attack). Each of these scenarios is tested under both static and dynamic conditions. These scenarios together make four general scenarios.

\textbf{Scenario Configurations:}
To rule out the possibility that the car reacts solely because of the camellia flower image or the presence of pedestrians and not the adversarial patch, we design several scenario configurations. For the single attack, we test three configurations:

\begin{enumerate}
    \item The T-shirt contains the adversarial patch (patch)
    \item The T-shirt shows only the camellia image, without any adversarial perturbation (camellia)
    \item The T-shirt is unchanged and contains no image (single-benign)
\end{enumerate}

\noindent For the collusion attack, we test seven configurations:

\begin{enumerate}
    \item No patches on either pedestrian (collusion-benign)
    \item Both halves of the adversarial patch are present (patch-both)
    \item Only the left half of the patch is present (patch-left)
    \item Only the right half of the patch is present (patch-right)
    \item Both halves of the camellia image are used instead of the patch (camellia-both)
    \item Only the left half of the camellia image is used (camellia-left).
    \item Only the right half of the camellia image is used (camellia-right).
\end{enumerate}

Testing each half separately allows us to assess whether only half of the patch retains effectiveness or whether the full pattern is required for a successful attack. Additionally, we run each scenario 10 times to account for non-determinism or possible run-time errors.

\subsection{Evaluation Metrics and Configurations}
To evaluate the generated patch, we measure the Attack Success Rate (ASR), a metric commonly used in adversarial patch research, at the system-level, and we measure the Stop-sign-To-Pedestrian (STP) ratio, defined as the number of stop-sign detections divided by the number of pedestrian detections, at the model-level.  

In the \textbf{model-level evaluation}, we examine the language model's output produced by the Simlingo agent frame by frame, counting how many times a stop sign is detected and how many times pedestrians are detected. We then compute STP and average it across  10 runs. A higher ratio indicates a stronger and cleaner execution of the attack. 

\begin{figure}
    \centering
    \includegraphics[width=0.65\columnwidth]{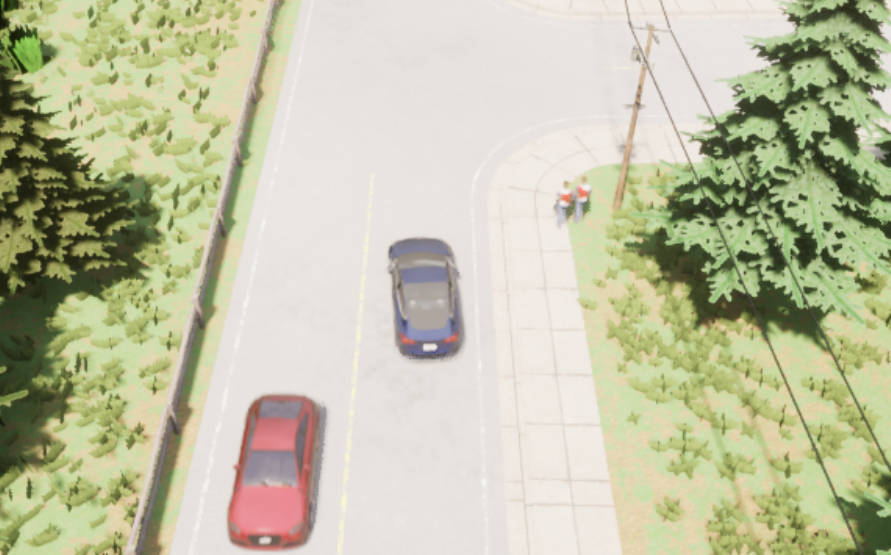}
    \caption[The car stopping at the intersection]{Example of a car coming to a full stop due to the adversarial pedestrians}
    \label{fig:system_example}
\end{figure}

In the \textbf{system-level evaluation}, an attack is considered successful only if it causes the car to come to a complete stop (see Fig.~\ref{fig:system_example}) before moving again, and the stopping is triggered by the detection of the stop sign rather than by pedestrians or any other objects. 
At the end of each simulation, we classify the attack as successful or not, and then we get ASR across 10 runs.

When comparing two attack scenarios with each other (e.g., single attack vs collusive attack), we use Fisher’s exact test~\cite{fisher1956mathematics} to assess the statistical significance of the difference between the ASR (resp. STP) values observed for the two attack scenarios at the system (resp. model) level. We also report the Odds Ratio (OR), which measures the strength and direction of the difference, indicating whether the proportions of successful vs unsuccessful system-level attacks (resp. stop sign vs pedestrian model-level detections) is different: OR $>$ 1 indicates that the first attack scenario has a higher proportion of successful vs unsuccessful attacks (resp. stop sign vs pedestrian detections) than the second; 0 $<$ OR $<$ 1, the opposite. 
We consider such a difference to be statistically significant when Fisher's $p$-value is lower than 0.05.

We conducted our experiments on a workstation equipped with an AMD Ryzen Threadripper PRO 5955WX CPU, an NVIDIA RTX 4000 GPU with 20GB GDDR6 memory, and running Ubuntu 22.04 as the operating system.
\section{Results}
\label{sec:results}

First of all, we evaluated the patch concealed within the camellia image on the surrogate model, YOLOv5. The patch on the camellia achieved ASR = 32.57\%, whereas the camellia image alone (without the patch) achieved only ASR = 0.66\%. This demonstrates that the adversarial patch is essential for the attack to succeed on the surrogate model, and that the camellia image is by itself ineffective.

After this initial validation on the surrogate model, below we answer the three research questions.



\begin{table}[ht]
\centering
\caption{ASR: number of successful attacks at the system-level over 10 runs}

\label{tab:results}

\begin{tabular}{|lcc|}

\hline
\rowcolor[HTML]{C0C0C0} 
\multicolumn{3}{|c|}{\cellcolor[HTML]{9B9B9B}Single}                                                                                      \\ \hline
\multicolumn{1}{|l|}{\textbf{Configuration}}                & \multicolumn{1}{c|}{\textbf{Dynamic}}                    & \textbf{Static} \\ \hline
\multicolumn{1}{|l|}{Benign}                                & \multicolumn{1}{c|}{0}                                   & 0               \\ \hline
\rowcolor[HTML]{EFEFEF} 
\multicolumn{1}{|l|}{\cellcolor[HTML]{EFEFEF}Camellia}      & \multicolumn{1}{c|}{\cellcolor[HTML]{EFEFEF}0}           & 0               \\ \hline
\multicolumn{1}{|l|}{Adversarial}                           & \multicolumn{1}{c|}{0}                                   & 0               \\ \hline
\rowcolor[HTML]{C0C0C0} 
\multicolumn{3}{|c|}{\cellcolor[HTML]{9B9B9B}Collusion}                                                                                   \\ \hline
\multicolumn{1}{|l|}{Benign}                                & \multicolumn{1}{c|}{0.1}                                  & 0               \\ \hline
\rowcolor[HTML]{EFEFEF} 
\multicolumn{1}{|l|}{\cellcolor[HTML]{EFEFEF}Patch-Both}    & \multicolumn{1}{c|}{\cellcolor[HTML]{EFEFEF}\textbf{0.5}} & 0               \\ \hline
\multicolumn{1}{|l|}{Patch-Left}                            & \multicolumn{1}{c|}{0.3}                                  & 0               \\ \hline
\rowcolor[HTML]{EFEFEF} 
\multicolumn{1}{|l|}{\cellcolor[HTML]{EFEFEF}Patch-Right}   & \multicolumn{1}{c|}{\cellcolor[HTML]{EFEFEF}0.2}          & 0               \\ \hline
\multicolumn{1}{|l|}{Camellia-Both}                         & \multicolumn{1}{c|}{0.3}                                  & 0               \\ \hline
\rowcolor[HTML]{EFEFEF} 
\multicolumn{1}{|l|}{\cellcolor[HTML]{EFEFEF}Camellia-Left} & \multicolumn{1}{c|}{\cellcolor[HTML]{EFEFEF}0.3}          & 0               \\ \hline
\multicolumn{1}{|l|}{Camellia-Right}                        & \multicolumn{1}{c|}{0.2}                                  & 0              
\\ \hline
\end{tabular}
\end{table}
\begin{table}[ht]
\centering
\caption{Average stop-sign and pedestrian detection from the Simlingo agent (i.e., model-level detection) in the single attack scenario over 10 runs}
\label{tab:single_ratio}
\begin{tabular}{lccc}
\rowcolor[HTML]{C0C0C0} 
\multicolumn{4}{c}{\cellcolor[HTML]{9B9B9B}Static}                                                                                                                                                                                                                        \\ \hline
\rowcolor[HTML]{FFFFFF} 
\multicolumn{1}{|l|}{\cellcolor[HTML]{FFFFFF}\textbf{Configuration}} & \multicolumn{1}{c|}{\cellcolor[HTML]{FFFFFF}\textbf{Pedestrians}} & \multicolumn{1}{c|}{\cellcolor[HTML]{FFFFFF}\textbf{Stop Signs}} & \multicolumn{1}{c|}{\cellcolor[HTML]{FFFFFF}\textbf{STP Ratio}} \\ \hline
\rowcolor[HTML]{EFEFEF} 
\multicolumn{1}{|l|}{\cellcolor[HTML]{EFEFEF}No-pedestrian}          & \multicolumn{1}{c|}{\cellcolor[HTML]{EFEFEF}0.4 $\pm$ 0.70}       & \multicolumn{1}{c|}{\cellcolor[HTML]{EFEFEF}6.00 $\pm$ 2.16}     & \multicolumn{1}{c|}{\cellcolor[HTML]{EFEFEF}15}             \\ \hline
\rowcolor[HTML]{FFFFFF} 
\multicolumn{1}{|l|}{\cellcolor[HTML]{FFFFFF}Benign}                 & \multicolumn{1}{c|}{\cellcolor[HTML]{FFFFFF}4.60 $\pm$ 2.55}      & \multicolumn{1}{c|}{\cellcolor[HTML]{FFFFFF}10.10 $\pm$ 2.02}    & \multicolumn{1}{c|}{\cellcolor[HTML]{FFFFFF}2.19}           \\ \hline
\rowcolor[HTML]{EFEFEF} 
\multicolumn{1}{|l|}{\cellcolor[HTML]{EFEFEF}Camellia}               & \multicolumn{1}{c|}{\cellcolor[HTML]{EFEFEF}3.30 $\pm$ 2.06}      & \multicolumn{1}{c|}{\cellcolor[HTML]{EFEFEF}19.20 $\pm$ 1.48}    & \multicolumn{1}{c|}{\cellcolor[HTML]{EFEFEF}5.81}           \\ \hline
\multicolumn{1}{|l|}{Patch}                                    & \multicolumn{1}{c|}{3.50 $\pm$ 2.01}                              & \multicolumn{1}{c|}{18.20 $\pm$ 2.10}                            & \multicolumn{1}{c|}{5.2}                                    \\ \hline
\rowcolor[HTML]{C0C0C0} 
\multicolumn{4}{c}{\cellcolor[HTML]{9B9B9B}Dynamic}                                                                                                                                                                                                                     \\ \hline
\multicolumn{1}{|l|}{Benign}                                         & \multicolumn{1}{c|}{5.50 $\pm$ 1.27}                              & \multicolumn{1}{c|}{7.00 $\pm$ 1.05}                             & \multicolumn{1}{c|}{1.27}                                   \\ \hline
\rowcolor[HTML]{EFEFEF} 
\multicolumn{1}{|l|}{\cellcolor[HTML]{EFEFEF}Camellia}               & \multicolumn{1}{c|}{\cellcolor[HTML]{EFEFEF}3.10 $\pm$ 1.52}      & \multicolumn{1}{c|}{\cellcolor[HTML]{EFEFEF}14.40 $\pm$ 4.12}    & \multicolumn{1}{c|}{\cellcolor[HTML]{EFEFEF}4.64}           \\ \hline
\multicolumn{1}{|l|}{Patch}                                    & \multicolumn{1}{c|}{4.20 $\pm$ 1.99}                              & \multicolumn{1}{c|}{11.40 $\pm$ 3.47}                            & \multicolumn{1}{c|}{2.71}                                   \\ \hline
\end{tabular}
\end{table}

\subsection{RQ1 [Single Attack]}

At the system-level, the number of successful single pedestrian attacks over 10 runs (i.e., ASR),is presented in Table~\ref{tab:results} (top).  
It shows that the single-pedestrian attack did not succeed in causing the car to stop due to a perceived stop sign; thus, the single attack is not effective.

To gain further insight into the model-level detection behavior in the single pedestrian attack scenario, we examine the agent’s average number of stop-sign and pedestrian detections, their standard deviations, and the STP ratio in Table~\ref{tab:single_ratio}. Table~\ref{tab:single_ratio} also reports the agent’s detection outputs when no pedestrians are in sight, providing a baseline.

The detection results indicate that the agent reports more stop signs than pedestrians in both static and dynamic scenarios when the camellia image or the adversarial patch is present. This suggests that the attack does mislead the agent to some extent, even though the system ultimately does not execute a full stop (see Table~\ref{tab:results} (top)).


Furthermore, in the no-pedestrian configuration, the agent still detects stop signs when approaching the intersection. This shows that the agent behaves cautiously near intersections, independently of any attack.

\begin{tcolorbox}[boxrule=0pt,frame hidden,sharp corners,enhanced,borderline north={1pt}{0pt}{black},borderline south={1pt}{0pt}{black},boxsep=2pt,left=2pt,right=2pt,top=2.5pt,bottom=2pt]
\textbf{Answer to RQ1}: The single-pedestrian attack is not effective at the system level: across all static and dynamic trials, the vehicle never stopped due to a perceived stop sign. At the model level, however, the agent consistently over-reports stop signs when the camellia or adversarial patch is present, indicating that the attack does induce some misclassification, but not strongly enough to trigger a full vehicle stop.
\end{tcolorbox}

\subsection{RQ2 [Collusion Attack]}


For the collusion scenarios, the number of successful attacks at the system-level over 10 runs is shown in Table~\ref{tab:results} (bottom). The collusion attack succeeded in the dynamic scenario (Patch-Both) with ASR = 50\%, when the two pedestrians moved alongside the car to remain within its field of vision. In contrast, the static scenario did not produce any successful attacks.

We also observe that although each half of the patch (Patch-Left and Patch-Right) was capable of stopping the car on its own, the Patch-Both configuration outperformed both individual halves, demonstrating the increased effectiveness of the complete patch. Additionally, the camellia image alone produced some successful stops, indicating that even the visual resemblance of the camellia to a stop sign can occasionally mislead the vehicle.


At the model-level, Table~\ref{tab:collusion_ratio} presents the agent’s detection outputs for the collusion scenarios, enabling direct comparison with the single-pedestrian results (Table~\ref{tab:single_ratio}). The agent identified stop signs more frequently than pedestrians in the dynamic collusion scenarios than in the single-pedestrian scenarios. The agent also detected more pedestrians overall in the collusion setup, reflecting the presence of two pedestrians instead of one. The highest STP values are always achieved in the dynamic collusive scenarios, where the camellia (both) and the patch (both) are respectively associated with STP = 25.55 and STP = 13.93. By contrast, in the single pedestrian scenario (see Table~\ref{tab:single_ratio}), the highest STP in the presence of pedestrians is 5.81 for camellia-static, which becomes 4.64 in the dynamic case.

Looking at Fisher’s test in Table~\ref{tab:fisher}, comparing the collusion attack to the single 
attack at the model level, we observe that the odds ratio is greater than 1 in the dynamic setting. This indicates that the odds of detecting a stop sign (rather than a pedestrian) are higher in the collusion attack than in the single attack, highlighting the increased effectiveness of the collusion approach, particularly in the dynamic scenario. Moreover, the consistently low $p$-values (all below 0.05) indicate that the difference between STP values is incompatible with the null hypothesis (stating that STP values are the same in the two attack scenarios), reinforcing the statistical reliability of the observed effect.

It should be noticed that we cannot provide a table similar to Table~\ref{tab:fisher} for the system-level results, because single pedestrian attacks have ASR = 0 (see Table~\ref{tab:single_ratio}, top), while dynamic collusive attacks have ASR $>$ 0 (see Table~\ref{tab:single_ratio}, bottom), which make the OR metric diverge. This is, of course, a very strong indication that at the system level, the difference between single and collusive attacks is huge.

\begin{table*}[ht]
\caption{Fisher’s exact test, comparing collusion and single-attack scenarios at the model level. Rows report OR and whether $p$-value \textless 0.05, thereby rejecting the null hypothesis. We use both parts of the patch in the collusion scenarios in this comparison.}
\centering
\label{tab:fisher}
\begin{tabular}{|l|ccc|ccc|}
\hline
\rowcolor[HTML]{9B9B9B} 
\multicolumn{1}{|c|}{\cellcolor[HTML]{9B9B9B}-}                                       & \multicolumn{3}{c|}{\cellcolor[HTML]{9B9B9B}Dynamic}                                                                                           & \multicolumn{3}{c|}{\cellcolor[HTML]{9B9B9B}Static}                                                                                            \\ \hline
\rowcolor[HTML]{EFEFEF} 
\multicolumn{1}{|c|}{\cellcolor[HTML]{FFFFFF}\textbf{-}} & \multicolumn{1}{c|}{\cellcolor[HTML]{EFEFEF}\textbf{Patch}} & \multicolumn{1}{c|}{\cellcolor[HTML]{EFEFEF}\textbf{Camellia}} & \textbf{Benign} & \multicolumn{1}{c|}{\cellcolor[HTML]{EFEFEF}\textbf{Patch}} & \multicolumn{1}{c|}{\cellcolor[HTML]{EFEFEF}\textbf{Camellia}} & \textbf{Benign} \\ \hline
\rowcolor[HTML]{FFFFFF} 
\textbf{Odds Ratio}                                                                   & \multicolumn{1}{c|}{\cellcolor[HTML]{FFFFFF}5.1349}         & \multicolumn{1}{c|}{\cellcolor[HTML]{FFFFFF}5.5015}            & 2.2205          & \multicolumn{1}{c|}{\cellcolor[HTML]{FFFFFF}0.5978}         & \multicolumn{1}{c|}{\cellcolor[HTML]{FFFFFF}0.6027}            & 0.5806          \\ \hline
\rowcolor[HTML]{FFFFFF} 
\textbf{P-value \textless 0.05}                                                       & \multicolumn{1}{c|}{\cellcolor[HTML]{FFFFFF}Yes}            & \multicolumn{1}{c|}{\cellcolor[HTML]{FFFFFF}Yes}               & Yes             & \multicolumn{1}{c|}{\cellcolor[HTML]{FFFFFF}Yes}            & \multicolumn{1}{c|}{\cellcolor[HTML]{FFFFFF}Yes}               & Yes             \\ \hline
\end{tabular}
\end{table*}
\begin{table}[ht]
\centering
\caption{Average stop-sign and pedestrian detection from the Simlingo agent (i.e., model-level) in the collusion attack scenario over 10 runs}
\label{tab:collusion_ratio}
\begin{tabular}{lccc}
\rowcolor[HTML]{C0C0C0} 
\multicolumn{4}{c}{\cellcolor[HTML]{9B9B9B}Static}                                                                                                                                                                                           \\ \hline
\multicolumn{1}{|l|}{\textbf{Configuration}}                  & \multicolumn{1}{c|}{\textbf{Pedestrians}}                  & \multicolumn{1}{c|}{\textbf{Stop Signs}}                   & \multicolumn{1}{c|}{\textbf{STP Ratio}}                \\ \hline
\rowcolor[HTML]{EFEFEF} 
\multicolumn{1}{|l|}{\cellcolor[HTML]{EFEFEF}Benign}          & \multicolumn{1}{c|}{\cellcolor[HTML]{EFEFEF}17.10 $\pm$ 4.63} & \multicolumn{1}{c|}{\cellcolor[HTML]{EFEFEF}21.80 $\pm$ 2.97} & \multicolumn{1}{c|}{\cellcolor[HTML]{EFEFEF}1.27}  \\ \hline
\multicolumn{1}{|l|}{Patch\_both}                             & \multicolumn{1}{c|}{8.30 $\pm$ 3.71}                          & \multicolumn{1}{c|}{25.80 $\pm$ 3.26}                         & \multicolumn{1}{c|}{3.10}                          \\ \hline
\rowcolor[HTML]{EFEFEF} 
\multicolumn{1}{|l|}{\cellcolor[HTML]{EFEFEF}Patch\_left}     & \multicolumn{1}{c|}{\cellcolor[HTML]{EFEFEF}13.80 $\pm$ 2.82} & \multicolumn{1}{c|}{\cellcolor[HTML]{EFEFEF}22.80 $\pm$ 3.74} & \multicolumn{1}{c|}{\cellcolor[HTML]{EFEFEF}1.65}  \\ \hline
\multicolumn{1}{|l|}{Patch\_right}                            & \multicolumn{1}{c|}{12.70 $\pm$ 4.67}                         & \multicolumn{1}{c|}{26.40 $\pm$ 3.31}                          & \multicolumn{1}{c|}{2.07}                          \\ \hline
\rowcolor[HTML]{EFEFEF} 
\multicolumn{1}{|l|}{\cellcolor[HTML]{EFEFEF}Camellia\_both}  & \multicolumn{1}{c|}{\cellcolor[HTML]{EFEFEF}7.50 $\pm$ 3.84}  & \multicolumn{1}{c|}{\cellcolor[HTML]{EFEFEF}26.30 $\pm$ 2.83} & \multicolumn{1}{c|}{\cellcolor[HTML]{EFEFEF}3.50}  \\ \hline
\multicolumn{1}{|l|}{Camellia\_left}                          & \multicolumn{1}{c|}{12.10 $\pm$ 5.24}                                      & \multicolumn{1}{l|}{23.10 $\pm$ 3.51}                                      & \multicolumn{1}{c|}{1.90}                              \\ \hline
\rowcolor[HTML]{EFEFEF} 
\multicolumn{1}{|l|}{\cellcolor[HTML]{EFEFEF}Camellia\_right} & \multicolumn{1}{c|}{\cellcolor[HTML]{EFEFEF}11.90 $\pm$ 4.75}              & \multicolumn{1}{c|}{\cellcolor[HTML]{EFEFEF}26.30 $\pm$ 3.40}              & \multicolumn{1}{c|}{\cellcolor[HTML]{EFEFEF}2.21}      \\ \hline
\rowcolor[HTML]{C0C0C0} 
\multicolumn{4}{c}{\cellcolor[HTML]{9B9B9B}Dynamic}                                                                                                                                                                                          \\ \hline
\multicolumn{1}{|l|}{Benign}                                  & \multicolumn{1}{c|}{4.60 $\pm$ 3.57}                          & \multicolumn{1}{c|}{13.00 $\pm$ 4.27}                         & \multicolumn{1}{c|}{2.82}                          \\ \hline
\rowcolor[HTML]{EFEFEF} 
\multicolumn{1}{|l|}{\cellcolor[HTML]{EFEFEF}Patch\_both}     & \multicolumn{1}{c|}{\cellcolor[HTML]{EFEFEF}1.60 $\pm$ 1.35}  & \multicolumn{1}{c|}{\cellcolor[HTML]{EFEFEF}22.30 $\pm$ 4.81} & \multicolumn{1}{c|}{\cellcolor[HTML]{EFEFEF}13.93} \\ \hline
\multicolumn{1}{|l|}{Patch\_left}                             & \multicolumn{1}{c|}{2.60 $\pm$ 2.32}                          & \multicolumn{1}{c|}{18.20 $\pm$ 4.26}                         & \multicolumn{1}{c|}{7}                             \\ \hline
\rowcolor[HTML]{EFEFEF} 
\multicolumn{1}{|l|}{\cellcolor[HTML]{EFEFEF}Patch\_right}    & \multicolumn{1}{c|}{\cellcolor[HTML]{EFEFEF}7.10 $\pm$ 5.00}  & \multicolumn{1}{c|}{\cellcolor[HTML]{EFEFEF}20.20 $\pm$ 3.91} & \multicolumn{1}{c|}{\cellcolor[HTML]{EFEFEF}2.84}  \\ \hline
\multicolumn{1}{|l|}{Camellia\_both}                          & \multicolumn{1}{c|}{0.90 $\pm$ 0.99}                          & \multicolumn{1}{c|}{23.00 $\pm$ 5.27}                         & \multicolumn{1}{c|}{25.55}                         \\ \hline
\rowcolor[HTML]{EFEFEF} 
\multicolumn{1}{|l|}{\cellcolor[HTML]{EFEFEF}Camellia\_left}  & \multicolumn{1}{c|}{\cellcolor[HTML]{EFEFEF}3.91 $\pm$ 2.74}  & \multicolumn{1}{c|}{\cellcolor[HTML]{EFEFEF}17.09 $\pm$ 7.44} & \multicolumn{1}{c|}{\cellcolor[HTML]{EFEFEF}4.37}  \\ \hline
\multicolumn{1}{|l|}{Camellia\_right}                         & \multicolumn{1}{c|}{8.70 $\pm$ 3.23}                          & \multicolumn{1}{c|}{19.80 $\pm$ 3.91}                         & \multicolumn{1}{c|}{2.27}                          \\ \hline
\end{tabular}
\end{table}

\begin{tcolorbox}[boxrule=0pt,frame hidden,sharp corners,enhanced,borderline north={1pt}{0pt}{black},borderline south={1pt}{0pt}{black},boxsep=2pt,left=2pt,right=2pt,top=2.5pt,bottom=2pt]
\textbf{Answer to RQ2}: The collusion attack is effective in the dynamic setting, achieving up to 50\% success rate, whereas the single attack never succeeds. Although each half of the patch can trigger a stop on its own, the full patch (both) configuration is stronger, confirming the benefit of distributing the patch across two pedestrians.
\end{tcolorbox}

\subsection{RQ3 [Dynamic Aspect]}

\begin{figure*}
    \centering
    \includegraphics[width=0.70\textwidth]{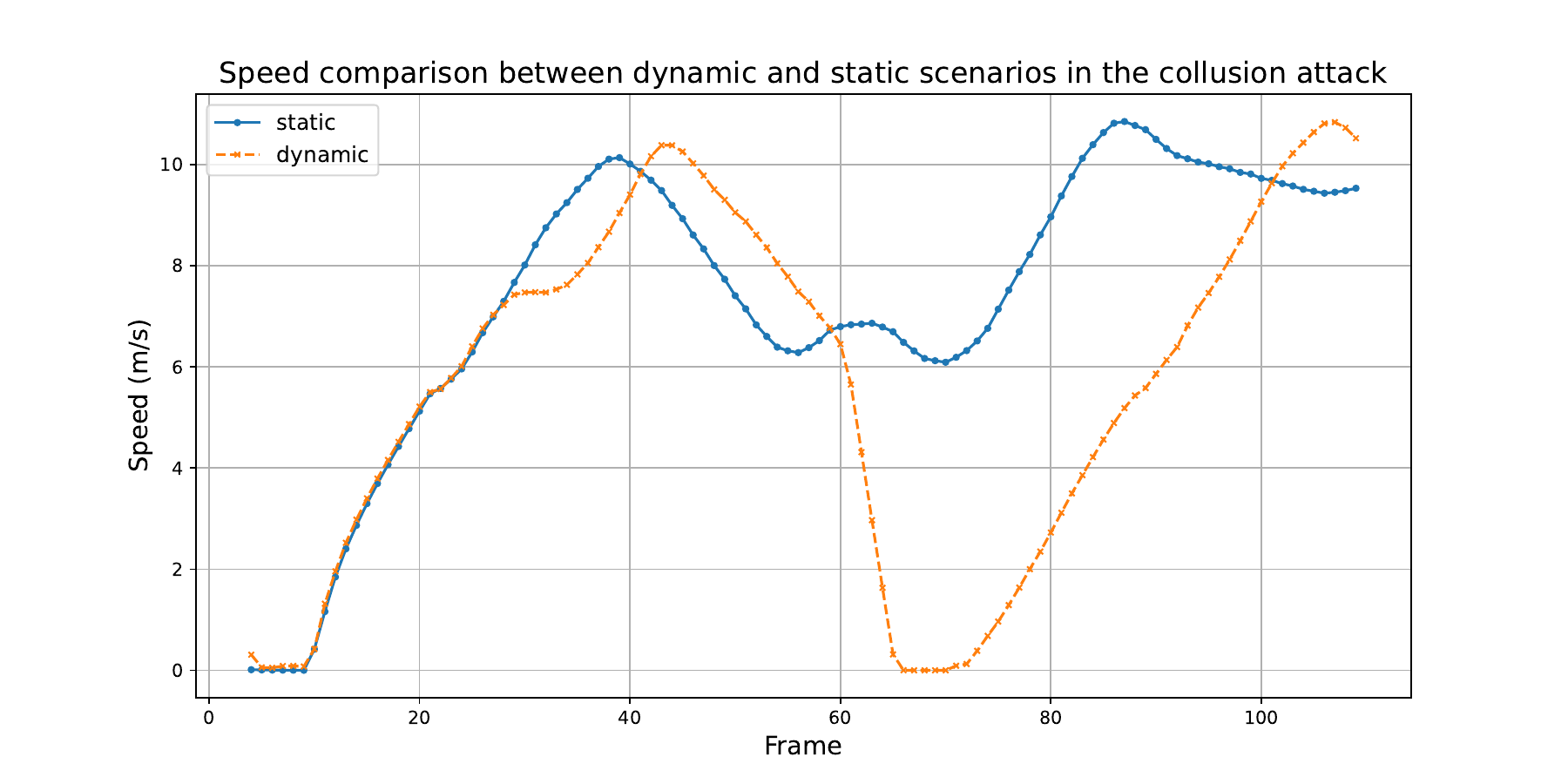}
    \caption[Speed comparison between dynamic and static scenarios in the collusion attack]{Speed comparison between dynamic and static scenarios in the collusion attack}
    \label{fig:speed}
\end{figure*}

Tables~\ref{tab:results}, \ref{tab:single_ratio}, and \ref{tab:collusion_ratio} indicate that the dynamic aspect is the key factor enabling the attacks to succeed. Even with two pedestrians maximizing the visible patch area, only the dynamic configuration caused the AC to stop. A plausible explanation is that continuous movement kept the patch within the camera’s field of view, producing a sustained sequence of speed-reduction cues, as depicted in Fig.~\ref{fig:speed}. These cues slowed the vehicle before it reached the intersection, making a full stop more likely once stop-signs are repeatedly detected. 

Examining the Simlingo outputs for the collusion scenarios in Table~\ref{tab:collusion_ratio}, we observe that in the dynamic setting, and across nearly all configurations, the agent reported fewer pedestrian detections than in the static setting. Additionally, the results of Fisher’s test in Table~\ref{tab:fisher} support this finding. When comparing the collusion attack to the single attack in the dynamic setting,  ORs are consistently greater than 1, indicating a strong effect of collusion. In contrast, the static setting shows ORs below 1, highlighting that the advantage of collusion emerges primarily in dynamic scenarios. These results suggest that pedestrians are more rarely recognized in the dynamic scenarios, further contributing to the attack’s success.


\begin{tcolorbox}[boxrule=0pt,frame hidden,sharp corners,enhanced,borderline north={1pt}{0pt}{black},borderline south={1pt}{0pt}{black},boxsep=2pt,left=2pt,right=2pt,top=2.5pt,bottom=2pt]
\textbf{Answer to RQ3}: Adding dynamism is crucial for making the attack effective: only the dynamic collusion scenario caused the vehicle to fully stop. Pedestrian movement keeps the patch visible longer, sustaining speed-reduction cues that ultimately trigger a full stop. Dynamic motion also helps hide the pedestrians, further strengthening the attack.
\end{tcolorbox}
\section{Discussion}
\label{sec:discussion}

\subsection{Pedestrians' Role}
Although the pedestrians were used solely as carriers of the adversarial patch, they indirectly assisted in making the car stop. Despite pedestrians being detected less frequently in the dynamic scenario, their presence introduced additional uncertainty into the agent’s decision-making, causing the car to behave more cautiously and reduce its speed, ultimately facilitating a complete stop. Importantly, we counted an attack as successful only when the agent reported detecting stop signs during the stopping event, not pedestrians. This ensures that the observed stop was attributable to the stop-sign signal induced by the adversarial patch, rather than to the presence of pedestrians.

\subsection{Transferability}
In the patch-generation phase, we used YOLOv5s and trained it without including Town07 in our dataset. Nevertheless, the results demonstrated that the attack transferred successfully to both a different model and previously unseen environments, such as Town07 in the CARLA simulator. 

\subsection{Camellia Image}
Looking across all results, we observe that even the camellia flower image, without any adversarial patch, was effective at both the model level and the system level. This suggests that when an image naturally resembles the target class (in our case, a stop sign), it can induce misclassification with surprising ease. Moreover, because our attack relies on placing the patch on pedestrians, it is highly deployable in practice: individuals with no expertise in adversarial patch generation could simply print a red camellia flower on a T‑shirt and unintentionally (or intentionally) cause an autonomous car to misclassify pedestrians as stop signs, ultimately forcing the vehicle to come to a complete stop.

\section{Threats to Vailidty}
\label{sec:threats}

\textbf{Internal validity threats}:
We addressed several potential threats to the internal validity of our results by designing specific scenario configurations. First, we tested the camellia flower image alone to eliminate the possibility that the adversarial patch was not contributing meaningfully to the attack. We also evaluated benign pedestrians wearing normal T-shirts to ensure that the vehicle’s behavior was caused by the printed images rather than merely by the presence of pedestrians. In the collusion attack, we further tested each half of the patch independently, confirming that using both halves together provides a stronger attack signal than using either half alone.

\textbf{Conclusion validity threats}:
Autonomous agents, in our case, the Simlingo agent, in the CARLA simulator, are not deterministic. Even when scenarios are configured identically (including weather and environmental conditions), the agents may not produce the same driving output in each frame across runs. This arises from small variations in object classification at a single frame, which can propagate to subsequent frames, ultimately leading to different agent outputs when examining the agent’s per-frame decisions. Although this variability does not significantly affect the agent’s overall ability to avoid obstacles, it introduces noise into our evaluation. To account for this non-determinism, we executed each scenario 10 times, and for the model-level evaluation, we reported the mean and standard deviation of the outputs. While running additional trials could further reduce uncertainty, ten repetitions provided results sufficiently stable for answering our research questions. Fisher's statistical test confirms that statistical significance could be achieved in our setting with 10 repetitions of each experiment.

\textbf{External validity threats}:
The generalizability of our findings to other AC simulators or to ACs operating in the real world is not ensured. To maximize the external validity of our results, we have chosen CARLA as the AC simulator, since it is widely used in research, it comes with a Leaderboard of high-performance driving agents, and it provides realistic driving environments. To support the replicability of our findings, possibly in different simulators or conditions, we provide our code as open source in our replication package (see link at the end of the paper).

\section{Conclusion \& Future Direction}
\label{sec:conclusion}
We introduced a novel attack on autonomous cars that leverages pedestrians as carriers of adversarial patches. By printing the patch directly onto a pedestrian’s T-shirt, the attack becomes extremely easy to deploy and can be executed simply by positioning pedestrians along the pavement. More importantly, pedestrians naturally offer mobility, allowing us to introduce dynamic adversarial behavior, an important and understudied dimension in attacks against ACs. We demonstrated that the simplicity of deployment enables attackers to use multiple pedestrians to increase both the spatial footprint and overall effectiveness of the attack.

Beyond hiding the adversarial pattern within a disguised image, pedestrians can also disable the attack at any moment by getting far from each other, as collusion was shown to be a key success factor of the proposed attack. This makes the attack not only effective but also highly covert. Moreover, pedestrians can adjust their walking speed to match that of the vehicle, maintaining the patch within the camera’s view for longer durations and further increasing the likelihood of system-level failure.
Together, these properties, the ease of deployment and the dynamic controllability, make this attack vector particularly dangerous for ACs operating in real-world environments.

An important future direction could be to investigate how existing or novel defence algorithms behave when confronted with dynamic or continuously visible adversarial patterns. Additionally, because several model-level results exhibited a high stop-sign-to-pedestrian ratio without causing the vehicle to stop, another promising direction is to analyse why certain attacks fail to propagate from model-level misclassification to system-level failure. Understanding this gap could provide helpful information on system-level robustness. Finally, our approach could be extended by exploring other traffic signs beyond stop signs and by evaluating its performance under diverse environmental conditions, such as varying weather, lighting, and scene complexity.

\section*{Replication Package}

The implementations, source code, and data are publicly available at: \\ https://github.com/MasoudJTehrani/DynamicDeception

\section*{Acknowledgment}
This work is funded by the European Union's Horizon Europe research and innovation programme under the project Sec4AI4Sec, grant agreement No 101120393.

\balance
\bibliographystyle{unsrt}  
\bibliography{bib}  

\end{document}